\documentclass[10pt,letterpaper]{article}
\usepackage{amsmath}
\usepackage{amssymb}
\usepackage{cite}
\usepackage{color}
\usepackage{opex3}
\usepackage{txfonts}
\usepackage{epstopdf}
\usepackage{amssymb}
\usepackage{tipa}
\usepackage{graphicx}
\usepackage{txfonts}
\usepackage{amssymb}
\usepackage{extarrows}

\begin{document}
\title{Compound photon blockade based on three mode system}
\author{Hongyu. Lin$^{1}$, Feng. Gao$^{1}$, Zhihai. Yao$^{2*}$ and Xiangyi Lou$^{1\dag}$}
\address{$^1$ College of Physics and electronic information, Baicheng Normal University, Baicheng - 137000, China\\
$^2$  Department of Physics, Changchun University of Science and Technology, Changchun - 130022, China \\
}
\email{*yaozh@cust.edu.cn, \dag luoxylgq@163.com}

\begin{abstract}
Based on the four wave mixing, a three mode nonlinear system is proposed. The single photon blockade is discussed through analytical analysis and numerical calculation. The analytical analysis shows that the conventional photon blockade and unconventional photon blockade can be realized at the same time, and the analytical conditions of the two kinds of blockade are the same. The numerical results show that the system not only has the maximum average photon number in the blockade region, but also can have strong photon anti-bunching in the region with small nonlinear coupling coefficient, which greatly reduces the experimental difficulty of the system. This optical system which can realize the compound photon blockade effect is helpful to realize the high-purity single photon source.
\end{abstract}






\section{Introduction}
In recent years, the realization of photon non-classical states is a research hotspot of quantum optics and quantum integration technology~\cite{01,02}, which is of great significance for the development and application of light quantum field in the future. In particular, the development of single photon source technology is the foundation for promoting the development of quantum communication~\cite{03}, quantum metrology~\cite{04}, and quantum information technology~\cite{05,06}. Therefore, the study of single photon source has become a focus of quantum physics. At present, there are two main ways to prepare single photon sources: one is weak light pulse attenuation, the other is parametric conversion based on nonlinear materials. The attenuation of weak light pulse has high multiphoton probability, which can not guarantee its single photon property. The preparation of single photon source according to parametric conversion requires that the system can only allow one photon to produce, or only the previous photon leaves the system, the next photon can enter or emit, that is, photon anti-bunching phenomenon. According to the different physical mechanisms of photon anti-bunching, it can be divided into two kinds. One is the photon anti-bunching phenomenon caused by the splitting of the energy level of the system depending on the strong second-order ($\chi^2$) or third-order ($\chi^3$) nonlinearity of the system. The photon anti-bunching phenomenon realized under this mechanism is called conventional photon blockaed (CPB). The other is unconventional photon blockade (UPB) based on quantum interference. In 1997, Imam$\bar{o}$glu et al. first observed the  photon blockaed effect in the experiment of optical cavity coupling and trapping atoms~\cite{07}. This discovery is also considered as a milestone in quantum optics and laser science.

The realization of CPB mainly depends on the quantum nonlinear resources of the system. There are two main ways to realize quantum nonlinear resources: one is to obtain them through high-order nonlinear effects, which requires the system to have high nonlinear electric polarizability to realize large nonlinear quantum systems, such as Kerr-nonlinear systems~\cite{08,09}. The second is the nonlinearity realized by the strong coupling between the optical microcavity and the atoms in the cavity, which is called quantum truncation~\cite{10,11,12}. In experiments, many schemes have been proposed for the realization of quantum nonlinear resources, such as optical force crystal~\cite{13}, superconducting circuit~\cite{14}, ultra cold atom capture~\cite{15,16}, etc. Most of its prototype systems are cavity quantum electrodynamics (QED) systems~\cite{17,18,19,20,21}. Birnbaum et al. realized the CPB in the optical microcavity for the first time~\cite{17}, and then Faraon et al. also found the CPB phenomenon in the system of embedding quantum dots in the resonant cavity of photonic crystals~\cite{23}. In addition, researchers have also predicted the existence of CPB in line QRD system~\cite{24}, optomechanical system~\cite{25,26,27}, waveguide system~\cite{28}. Ester et al. realized single photon emission in InGaAs / GaAs quantum dot system, and the second-order correlation function is only 0.02~\cite{29}. According to the research, when the second-order correlation function of the single photon emission system is less than 0.07, the efficiency of quantum computing can reach $0.9$, so the system reported by Ester et al. Has almost realized a perfect single photon source.

Unlike CPB, Liew and savona found a new mechanism to realize PB, in which the realization of photon anti-bunching does not depend on the nonlinear intensity of the system~\cite{30}. Subsequently, Carmichael and Bamba analyzed analytically that the physical mechanism behind the PB in weakly nonlinear systems is destructive quantum interference~\cite{31,32}. This kind of PB based on destructive quantum interference is called UPB. Flayac et al. discussed the input-output theory of UPB in a coupled two cavity system~\cite{33}. In the same year, Zhou et al. realized UPB in a typical two-mode system, and discussed the implementation conditions and parameter settings in detail~\cite{34}. With the development of research, UPB has also been realized in many similar weakly nonlinear coupled two mode or three mode systems~\cite{35,36,37,38,39}.

Recently, it has also been found that destructive quantum interference can also be used to improve the intensity of CPB, so as to improve the purity of single photon source~\cite{40}. In addition, researchers also found that CPB and UPB can not only coexist in the same quantum system, but also have the same coefficient region, which helps us to further understand the physical mechanism behind CPB and UPB~\cite{41}. Generally, the average number of single photons produced by CPB is high, but the purity is low, and the nonlinearity of the system is high. When UPB is realized, the value of the general second-order correlation function is very low, but the brightness is relatively poor. Valle et al. found that the advantages of UPB and CPB can be combined in the form of double drive~\cite{42}.

In this work, we propose a three mode nonlinear coupled system based on four wave mixing, and discuss the single photon blockade in the system. The optimal analytical solution of the system to realize CPB and UPB is obtained through analytical calculation. According to the calculation results, the system can not only realize CPB and UPB at the same time, but also have the same analytical conditions. This shows that in the region of the optimal analytical solution, the system can realize two kinds of photon blockade effects at the same time, which will help the system produce more severe single photon blockade effect, and provide the possibility to realize the single photon source with high purity. Through numerical analysis, it can be seen that there is the largest number of photons in the region where strong photon anti-bunching is realized, and through the discussion of nonlinear coupling coefficient $g$, photon blockade can also be realized in the smaller nonlinear coupling region, and the photon blockade becomes more severe with the increase of coupling coefficient, which also reduces the difficulty of experimental implementation of the system to a great extent.

The manuscript is organized as follows:
In Sec.~{\rm II}, we introduce the physical model and implementation of PB.
In Sec.~{\rm III}, we illustrate the numerical results for the photon blockad.
Conclusions are given in Sec.~{\rm IV}.

\section{Physical model and implementation of PB}
\label{sec:2}
Here, we consider a general three mode optical system, where the interaction between modes is four wave mixing. The Hamiltonian of the system can be expressed as
\begin{eqnarray}
\hat{H}&=&\omega_a\hat{a}^{\dag}\hat{a}\
+\omega_b\hat{b}^{\dag}\hat{b}+\omega_c\hat{c}^{\dag}\hat{c}
+g(\hat{a}^2\hat{b}^{\dag}\hat{c}^{\dag}+\hat{a}^{\dag2}\hat{b}\hat{c})
+F_a(\hat{a}^{\dag}e^{-i\omega_l}+\hat{a}e^{i\omega_l}),
\label{01}
\end{eqnarray}
The $\hat{a}(\hat{a}^{\dag})$, $\hat{b}(\hat{b}^{\dag})$ and $\hat{c}(\hat{c}^{\dag})$ denotes the annihilation (creation) operator of mode $a$ ,$b$ and $c$ respectively. The $g$ expresse the coefficient of nonlinear interactions. $F_a$ represents the driving coefficient of mode $a$ photons, and the driving frequency is $\omega_l$.
By using rotating operator $\hat{U}(t)=e^{i t(\omega_{l}\hat{a}^{\dag}\hat{a}+\omega_{l_1}\hat{b}^{\dag}\hat{b}\omega_{l_2}\hat{c}^{\dag}\hat{c})}$, we can get effective Hamiltonian of the system as
\begin{eqnarray}
\hat{H}_{eff}&=&\Delta_a\hat{a}^{\dag}\hat{a}\
+\Delta_b\hat{b}^{\dag}\hat{b}+\Delta_c\hat{c}^{\dag}\hat{c}+g(\hat{a}^2\hat{b}^{\dag}\hat{c}^{\dag}+\hat{a}^{\dag2}\hat{b}\hat{c})+F_a(\hat{a}^{\dag}+\hat{a}),
\label{02}
\end{eqnarray}
Here, the detuning amount $\Delta_a=\omega_a-\omega_{l}$, $\Delta_b=\omega_b-\omega_{l_1}$ and $\Delta_c=\omega_c-\omega_{l_2}$ and satisfies the relationship of $\Delta_b + \Delta_c = 2 \Delta_a$.
The Fock-state basis of the system is denoted by $|mnp\rangle$ with the
number $m$ denoting the photon number in mode $a$, $n$ denoting the photon number in mode $b$, and $p$ denoting the photon number in mode $c$.
Under the weak driving limit, we restrict the system containing a single photon in the mode $a$.

\subsection{Analytical analysis of conventional photon blockade in the system}
\begin{figure}[h]
\centering
\includegraphics[scale=0.60]{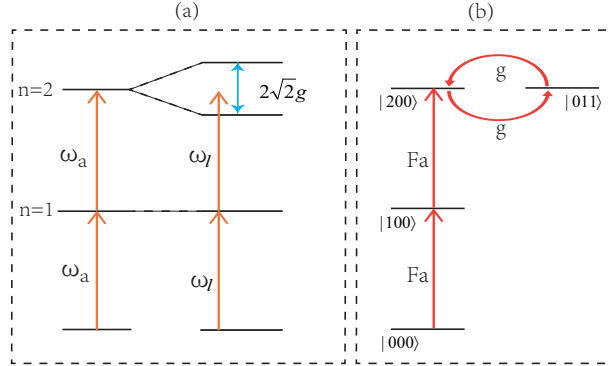}
\caption{(Color online) (a)Energy level diagram of conventional photon blockade realized by the system.(b)The photon state and interference path of non-traditional photon blockade are realized in the system.
} \label{fig1}
\end{figure}
According to the single excitation resonance condition, once the system resonantly absorbs a photon with frequency $\omega_{l}$, it can make it from photon state $|0-\rangle$ to photon state$|1-\rangle$. If the system has strong nonlinearity and leads to detuning between photon states $|1-\rangle$ and $|2-\rangle$, the system can no longer absorb a photon with the same frequency to activate photon state $|2-\rangle$. Therefore, only one existing photon in the system is radiated, the next photon can enter, and the photon anti-bunching phenomenon of photon intensity occurs, so as to realize single photon blockade. As shown in Fig.~\ref{fig1}.(a), if the current system has strong nonlinear coupling strength and single-mode driving frequency $\omega_{l}=\omega_{a}$, the system will not occupy photon state $|2-\rangle$ and realize single-photon blockade. Therefore, the condition of realizing conventional photon blockade in this system is $\Delta_a=0$. The splitting height of energy level 2 can be obtained by solving the eigenvalue of the system. Considering the photon blockade in mode a, we select $|200\rangle$ and $|011\rangle$ to form a closed space, the Hamiltonian can be expand with $|200\rangle$ and $|011\rangle$, which can be described as a matrix form
\begin{eqnarray}
\tilde{H}=
\begin{bmatrix}
2\Delta_a & \sqrt{2}g\\
 \sqrt{2}g & \Delta_b+\Delta_c
\end{bmatrix}, \label{03}
\end{eqnarray}
In fact, the occurrence of photon blockade is to convert the Poisson statistical photons in the input system into sub-Poisson statistical photons output. When the system realizes photon blockade, it will generally weaken with the increase of the driving light field intensity, so the system must meet the weak driving conditions in order to realize photon blockade, so we have neglect the driving terms. According to the coupling relationship of photons in each mode and the analytical conditions for realizing photon blockade in the system, by analyzing the above matrix we can get the eigenfrequencies $\omega^{(1)}_{+}$ and $\omega^{(1)}_{-}$, which can be written as
\begin{eqnarray}
\omega^{(1)}_{\pm}=\pm\sqrt{2}g.
\label{04}
\end{eqnarray}
Therefore, when the system realizes photon blockade, the splitting height of energy level 2 is $2\sqrt{2}g$.

\subsection{Analytical analysis of unconventional photon blockade in the system}
Different from the conventional photon blockade implementation, the nonlinearity of the system is much greater than the dissipation rate of the system. The implementation of unconventional photon blockade is based on destructive quantum interference between different paths, that is, there is no less than one path in the system from photon state $|1-\rangle$ to photon state $|2-\rangle$, and interference cancellation will occur between different paths that can reach photon state $|2-\rangle$, Thus, there can only be single photon states in the system. The energy-level structure and transition paths are shown in Fig .1.(b), there are two paths for the system to reach the two-photon state of mode $a$:
(i) $|000\rangle\stackrel{\underrightarrow{F_a}}{}|100\rangle\stackrel{\underrightarrow{F_a}}{}|200\rangle$.
(ii)$|000\rangle\stackrel{\underrightarrow{F_a}}{}|100\rangle\stackrel{\underrightarrow{F_a}}{}|200\rangle\stackrel{\underrightarrow{g}}{}|011\rangle\stackrel{\underrightarrow{g}}{}|200\rangle$.
When the optimal conditions for photon anti-bunching are satisfied, the photons come from different pathways destructive interference and the photons cannot occupy the state $|200\rangle$, that is, the UPB will occur. According to the system Hamiltonian, the wave function can be written as
\begin{eqnarray}
|\psi\rangle &=& C_{000}|000\rangle+C_{100}|100\rangle+C_{200}|200\rangle+C_{011}|011\rangle.
\label{05}
\end{eqnarray}
This is a good approximation when we restrict the system containing a single photon with mode $a$. Consider environmental influence we can treat the system by the non-Hermitian Hamiltonian
\begin{eqnarray}
\widetilde{H}=\hat{H}_{eff}-i\frac{\kappa}{2}\hat{a}^\dag\hat{a}
-i\frac{\kappa}{2}\hat{b}^\dag\hat{b}-i\frac{\kappa}{2}\hat{c}^\dag\hat{c}.
\label{06}
\end{eqnarray}
The $\hat{H}_{eff}$ is given in Eq.~(\ref{02}).
Substituting the wave function shown in Eq.~(\ref{05})
and non-Hermitian Hamiltonian into the Schr\"odinger equation
$i\partial_t|\psi\rangle=\widetilde{H}|\psi\rangle$,
we can obtain the coupled equations for the coefficients
\begin{eqnarray}
&&i\dot{C}_{000}=F_a C_{100},\nonumber\\
&&i\dot{C}_{100}=\sqrt{2}F_a C_{000}+(\Delta_a-\frac{i\kappa}{2})C_{100}+\sqrt{2}F_a C_{200},\nonumber\\
&&i\dot{C}_{200}=\sqrt{2}F_a C_{100}+2(\Delta_a-\frac{i\kappa}{2})C_{200}+\sqrt{2}g C_{011},\nonumber\\
&&i\dot{C}_{011}=\sqrt{2}g C_{200}+(\Delta_b-\frac{i\kappa}{2})C_{011}+(\Delta_b-\frac{i\kappa}{2})C_{011}.
\label{07}
\end{eqnarray}
Therefore, the steady-state coefficient equation can be expressed as
\begin{eqnarray}
&&F_a C_{100}=0,\nonumber\\
&&\sqrt{2}F_a C_{000}+(\Delta_a-\frac{i\kappa}{2})C_{100}+\sqrt{2}F_a C_{200}=0,\nonumber\\
&&\sqrt{2}F_a C_{100}+2(\Delta_a-\frac{i\kappa}{2})C_{200}+\sqrt{2}g C_{011}=0,\nonumber\\
&&\sqrt{2}g C_{200}+(\Delta_b-\frac{i\kappa}{2})C_{011}+(\Delta_b-\frac{i\kappa}{2})C_{011}=0.
\label{8}
\end{eqnarray}
Here, the probability amplitude always satisfies the relationship of $|C_{000}|\gg |C_{100}|\gg |C_{200}|$. Therefore, for the convenience of calculation, $|C_{200}|=0$, $|C_{000}|=1$ can be made, and the detuning of each photon mode satisfies $\Delta_a=\Delta_b+\Delta_c$. According to the above conditions, we can obtain the solution of equation Eq.~(\ref{8}), that is,
 \begin{eqnarray}
&&\Delta_b=-\Delta_c,
\label{9}
\end{eqnarray}
According to the nonlinear coupling relationship of the system, Eq.~(\ref{9}) can also be expressed as
\begin{eqnarray}
&&\Delta_a=0,
\label{10}
\end{eqnarray}
which is the optimal conditions for UPB in mode $a$.

Therefore, it can be seen from the above analysis results that the current system has the same analysis conditions when realizing CPB and UPB. In other words, CPB and UPB can be realized in the same blockade region, which can not only improve the intensity of CPB through destructive quantum interference, but also realize two kinds of photon blockade at the same time in a small nonlinear coupling region, so as to improve the purity of single photon source.

\section{The numerical results for the photon blockade}
\label{sec:3}
\begin{figure}[h]
\centering
\includegraphics[scale=0.60]{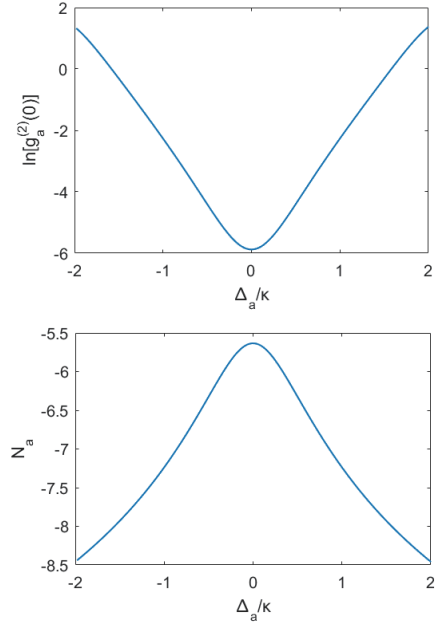}
\caption{(Color online) (a)The Logarithmic plot of the $g^{(2)}(0)$ as a function of $\Delta_a/\kappa$ for $F_a/\kappa=0.01$, $g/\kappa=3$ and $\Delta_b/\kappa=-\Delta_c/\kappa=1$. (b) The Logarithmic plot of the average photon number $N_a$ as a function of $\Delta_a/\kappa$ for $F_a/\kappa=0.01$, $g/\kappa=3$ and $\Delta_b/\kappa=-\Delta_c/\kappa=1$. }
\label{fig3}
\end{figure}
\begin{figure}[h]
\centering
\includegraphics[scale=0.60]{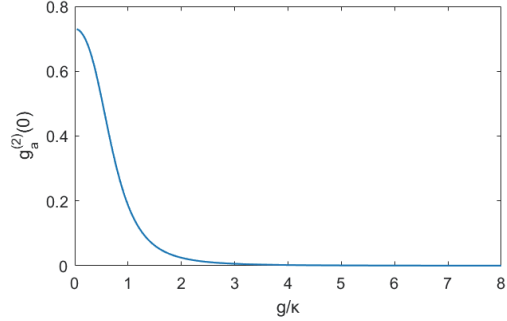}
\caption{(Color online) We plot of the $g^{(2)}(0)$ as a function of $g/\kappa$, with $F_a/\kappa=0.01$, $\Delta_a/\kappa=0$ and $\Delta_b/\kappa=-\Delta_c/\kappa=2$.}
\label{fig4}
\end{figure}
In the non-classical field theory, we can judge the statistical characteristics of photons by numerically solving the value of the second-order correlation function $g^{(2)}(0)$ of the system. If $g^{(2)}(0)<1$ , it indicates that photons in the system are sub Poisson distribution, that is, photons are anti-bunching, and strong photon anti-bunching indicates that single photon blockade occurs in the system. And the expression of $g^{(2)}(0)$ can be obtained by solving the master equation. The dynamics of the density matrix $\hat{\rho}$ of the system is governed by
\begin{eqnarray}
\frac{\partial\hat{\rho}}{\partial t}&=&-i[\hat{H},\rho]+\frac{\kappa_a}{2}(2\hat{a}\hat{\rho}\hat{a}^\dag+\frac{1}{2}\hat{a}^\dag\hat{a}\hat{\rho}+\frac{1}{2}\hat{\rho}\hat{a}^\dag\hat{a})\nonumber\\
&&+\frac{\kappa_b}{2}(2\hat{b}\hat{\rho}\hat{b}^\dag+\frac{1}{2}\hat{b}^\dag\hat{b}\hat{\rho}+\frac{1}{2}\hat{\rho}\hat{b}^\dag\hat{b})\nonumber\\
&&+\frac{\kappa_c}{2}(2\hat{c}\hat{\rho}\hat{c}^\dag+\frac{1}{2}\hat{c}^\dag\hat{c}\hat{\rho}+\frac{1}{2}\hat{\rho}\hat{c}^\dag\hat{c}).
\label{08}
\end{eqnarray}
The $\kappa_a$, $\kappa_b$ and $\kappa_c$ denote the decay rates of cavities $a$, $b$ and $c$, respectively. In order not to lose the general principle, we usually make the detuning satisfy the $\kappa_a=\kappa_b=\kappa_c=\kappa$ relationship. Here, we only need to consider $g^{(2)}(0)$ in the steady state. Therefore, we only need to set $\partial\hat{\rho}/\partial t=0$ to solve the master equation of the steady-state density operator $\hat{\rho}_s$. In this scheme, we will analyze the PB in $a$ mode. The statistical characteristics of photons will be described by the zero delay time correlation function, which is composed of
\begin{eqnarray}
g_{a}^{(2)}(0)=\frac{\langle
\hat{a}^{\dag}\hat{a}^{\dag}\hat{a}~\hat{a}\rangle}
{\langle \hat{a}^{\dag}\hat{a}\rangle^2}. \label{09}
\end{eqnarray}
The value of $g^{(2)}(0)$ can be calculated by numerically solving the master equation. If $g^{(2)}(0)<1$, it indicates that PB occurs, and a smaller $g^{(2)}(0)$ value indicates that a more severe PB effect occurs in the system, and the higher the purity of the single light source realized by the system. Therefore, making the system obtain smaller $g^{(2)}(0)$ value is an important condition to realize the ideal single photon source.

In Fig.~\ref{fig3}(a), we Logarithmic plot of the $g^{(2)}(0)$ as a function of $\Delta_a/\kappa$ for $F_a/\kappa=0.01$, $g/\kappa=3$ and $\Delta_b/\kappa=-\Delta_c/\kappa=1$. According to the calculation results, when the nonlinear coupling coefficient $g = 3$, the valley bottom value of the curve can reach $10^6$ where the analytical conditions are met, which shows that the system emits strong photon anti-bunching, and the numerical solution is consistent with the analytical solution. In Fig.~\ref{fig3}(b), we plot of the average photon number $N_a$ as a function of $\Delta_a/\kappa$ for $F_a/\kappa=0.01$, $g/\kappa=3$ and $\Delta_b/\kappa=-\Delta_c/\kappa=1$. According to the results in the figure, the region where photon blockade occurs has the largest average number of photons, which helps the system to be an alternative system for realizing a single photon source.

In Fig.~\ref{fig4},we plot of the $g^{(2)}(0)$ as a function of $g/\kappa$, with $F_a/\kappa=0.01$, $\Delta_a/\kappa=0$ and $\Delta_b/\kappa=-\Delta_c/\kappa=2$. According to the results in the figure, the current system can have strong single photon blocking in almost the whole value region of $g/\kappa$. according to the physical mechanism realized by CPB, the system needs to meet large nonlinearity ($g/\kappa\gg1$) to ensure its energy level splitting, while the results in the figure show that photon blockade can still be realized in the region of $g/\kappa<1$, This further verifies that the current system can realize two kinds of photon blockade at the same time, which is consistent with the analytical results.

\section{Conclusions}
\label{sec:4}
In this work, we propose a three mode nonlinear system model which can realize CPB and UPB simultaneously based on four wave mixing. The analytical calculation shows that the analytical conditions of the system are the same when realizing CPB and UPB, that is, the system can realize two kinds of photon blockade at the same coefficient interval, which will help to realize a single photon source with higher purity. According to the numerical calculation, the system can realize strong photon anti-bunching, and the numerical solution is consistent with the analytical solution, and has the maximum average photon number in the blockade region. Through the discussion of the nonlinear coupling coefficient $g$, it can be seen that the system can not only realize photon blockade in the area with small coupling coefficient, but also significantly enhance the photon blockade effect with the increase of coupling coefficient, which further verifies that the system can realize two kinds of photon blockade at the same time.

\section*{Acknowledgment}
This work is supported by the National Natural Science Foundation of China with Grants No. 11647054, the Science and Technology Development Program of Jilin province, China with Grant No. 2018-0520165JH.

\section*{Disclosures:} The authors declare no conflicts of interest.\\


\begin{thebibliography}{9}

\bibitem{01} L Davidovich, ``Sub-Poissonian processes in quantum optics", Rev. Mod. Phys \textbf{68}(1), 127-173 (1996)
\bibitem{02} A J Shields, ``Semiconductor quantum light sources", Nat. Photon \textbf{1}(4), 215-223 (2007).
\bibitem{03} V Scarani, H Bechmann-Pasquinucci, N J Cerf, M Dusek, N Lutkenhaus and M Peev, ``The security of practical quantum key distribution", Rev. Mod. Phys \textbf{81}(3), 1301-1350 (2009).
\bibitem{04} V Giovannetti, S Lloyd and L Maccone, ``Advances in quantum metrology", Nat. Photonics \textbf{5}, 222-229 (2011).
\bibitem{05} E Knill, R Laflamme and G J Milburn, ``A scheme for efficient quantum computation with linear optics", Nature \textbf{409}(6816), 46-52 (2001).
\bibitem{06} P Kok, W J Munro, K Nemoto, T C Ralph, J P Dowling and G J Milburn, ``Linear optical quantum computing with photonic qubits", Rev. Mod. Phys \textbf{79}(135), 135-174 (2007).
\bibitem{07} A Imamoglu, H Schmidt, G Woods and M Deutsch, ``Strongly Interacting Photons in a Nonlinear Cavity", Phys. Rev. Lett \textbf{79}, 1467-1470 (1997).
\bibitem{08} D Gerace and V Savona, ``Unconventional photon blockade in doubly resonant microcavities with second-order nonlinearity", Phys. Rev. A \textbf{89}, 031803(R) (2014).
\bibitem{09} H T Tan, ``Deterministic quantum superpositions and Fock states of mechanical oscillators via quantum interference in single-photon cavity optomechanics", Phys. Rev. A. \textbf{89}, 053829 (2014).
\bibitem{10}A Miranowicz, M Paprzycka, Y X Liu, J Bajer and F Nori, ``Two-photon and three-photon blockades in driven nonlinear systems", Phys. Rev. A \textbf{87}, 023809 (2013).
\bibitem{11}W Leonski and R Tanas, ``Possibility of producing the one-photon state in a kicked cavity with a nonlinear Kerr medium", Phys. Rev. A \textbf{49}, R20-R23 (1994).
\bibitem{12} G H Hovsepyan, A R Shahinyn and G Yu Kryuchkyan, ``Multiphoton blockades in pulsed regimes beyond stationary limits", Phys. Rev. A \textbf{90}, 013839 (2014).
\bibitem{13}M Eichenfield, J Chan, R M Camacho, et al, ``Optomechanical crystals", Nature \textbf{462}(7269),78-82 (2009).
\bibitem{14}D Teufel, T Donner, D Li, et al, ``Sideband cooling of micromechanical motion to the quantum ground state",  Nature \textbf{475}(7356), 359-363 (2011).
\bibitem{15}F Brennecke, S Ritter, T Donner, et al, ``Cavity Optomechanics with a Bose-Einstein Condensate", Science \textbf{322}(5899), 235-238 (2008).
\bibitem{16} S Gupta, K L Moore, K W Murch, et al, ``Cavity Nonlinear Optics at Low Photon Numbers from Collective Atomic Motion", Phys. Rev. Lett \textbf{99}(21), 213601 (2007).
\bibitem{17} K M Birnbaum, A Boca, R Miller, et al, ``Photon blockade in an optical cavity with one trapped atom", Nature \textbf{436}(7047), 87-90 (2005).
\bibitem{18}P R Berman, ``Zero eigenvalues of a photon blockade induced by a non-Hermitian Hamiltonian with a gain cavity", Phys. Rev. A \textbf{97}, 043819 (2018).
\bibitem{19}L Tian and H J Carmichael, `` Quantum trajectory simulations of two-state behavior in an optical cavity containing one atom", Phys. Rev. A \textbf{46}(11), R6801 (1992).
\bibitem{20}M J Werner and A Imamoglu, ``Photon-photon interactions in cavity electromagnetically induced transparency", Phys. Rev. A \textbf{61}(01), 011801R (1999).
\bibitem{21} R J Brecha, P R Rice and M Xiao, ``N two-level atoms in a driven optical cavity: Quantum dynamics of forward photon scattering for weak incident fifields", Phys. Rev. A \textbf{59},  2392 (1999).
\bibitem{23}A Faraon, I Fushman, D Englund, N Stoltz, P Petroff and J Vuckovic, ``Coherent generation of non-classical light on a chip via photon-induced tunnelling and blockade", Nat. Physics \textbf{4}(11), 859-863 (2008).
\bibitem{24} A J Hoffman, S J Srinivasan, S Schmidt, et al, ``Dispersive Photon Blockade in a Superconducting Circuit", Phys. Rev. Lett \textbf{107}(5), 053602 (2011).
\bibitem{25} J Q Liao and F Nori, ``Photon blockade in quadratically coupled optomechanical systems", Phys. Rev. A \textbf{88}, 023853 (2013).
 \bibitem{26} P Rabl, ``Photon Blockade Effffect in Optomechanical Systems", Phys. Rev. Lett \textbf{107}, 063601 (2011).
\bibitem{27}A Nunnenkamp, K B{\o}rkje and S M Girvinl, ``Single-Photon Optomechanics", Phys. Rev. Lett \textbf{107}, 063602 (2011).
\bibitem{28}C Lang, D Bozyigit, C Eichler, et al, ``Observation of Resonant Photon Blockade at Microwave Frequencies Using Correlation Function Measurements", Phys. Rev. Lett \textbf{106}(24), 243601 (2011).
\bibitem{29} P Este, L Lackmann, et al, ``Single photon emission based on coherent state preparation", Applied Physics Letters \textbf{91}(11), 111110 (2007).
\bibitem{30}T C H Liew and V Savona, ``Single Photons from Coupled Quantum Modes", Phys. Rev. Let \textbf{104}(18), 183601 (2010).
\bibitem{31}H J Carmichael, ``Photon Antibunching and Squeezing for a Single Atom in a Resonant Cavity", Phys. Rev. Lett \textbf{55}, 2790 (1985).
\bibitem{32}M Bamba, A Imamo?lu, I Carusotto and C Ciuti, ``Origin of strong photon antibunching in weakly nonlinear photonic molecules ", Phys. Rev. A \textbf{83}, 021802(R) (2011).
\bibitem{33}H Flayac and V Savona, ``Input-output theory of the unconventional photon blockade", Phys. Rev. A \textbf{88}(3), 033836 (2013).
\bibitem{34}Y H Zhou, H Z Shen and X X Yi, ``Unconventional photon blockade with second-order nonlinearity", Phys. Rev. A \textbf{92}(2), 023838 (2015).
\bibitem{35}H Flayac and V Savona, ``Unconventional photon blockade", Phys. Rev. A \textbf{96}(5), 053810 (2017).
\bibitem{36}E Z Casalengua, J C L Carreno, F P Laussy and E del Valle, ``Photon Statistics: Conventional and Unconventional Photon Statistics",  Laser Photon. Rev \textbf{14}, 1900279 (2020).
\bibitem{37} S Bijita and A K Sarma, ``Unconventional photon blockade in three-mode optomechanics", Phys. Rev. A \textbf{98}(1), 013826 (2018).
\bibitem{38}H Flayac, D Gerace and V Savona, ``An all-silicon single-photon source by unconventional photon blockade", Sci. Rep \textbf{5}(1), 11223 (2015).
\bibitem{39}H Z Shen, S Xu, Y H Zhou and G C Wang, ``Unconventional photon blockade from bimodal driving and dissipations in coupled semiconductor microcavities", J. Phys. B: At. Mol. Opt. Phys \textbf{51}(3), 035503 (2018).
\bibitem{40}J Tang, W Geng and X Xu, ``Quantum interference induced photon blockade in a coupled single quantum dot-cavity system", Sci. Rep \textbf{5}, 9252 (2015).
\bibitem{41}X Y Liang, Z L Duan, Q Guo, C J Liu, S G Guan and Y Ren, ``Antibunching effect of photons in a two-level emitter-cavity system", Phys. Rev. A \textbf{100}, 063834 (2019).
\bibitem{42}E Z Casalengua, J C L Carreno, F P Laussy and E del Valle, ``Photon Statistics: Conventional and Unconventional Photon Statistics", Laser Photon. Rev \textbf{14}, 1900279 (20209).
%
%
%
%
%
%
%
%
%
%
%

\end{thebibliography}
\end{document}